\documentclass{emulateapj}

\newcommand{\kms}{km~s$^{-1}$\,}
\newcommand{\masyr}{mas~yr$^{-1}$\,}
\newcommand{\msun}{${\cal M}_\odot$\,}

\begin{document}

\renewcommand{\topfraction}{1.0}
\renewcommand{\bottomfraction}{1.0}
\renewcommand{\textfraction}{0.0}

\title{Inner and Outer Orbits in 13 Resolved Hierarchical Stellar Systems}


\author{A. Tokovinin}
\affil{Cerro Tololo Inter-American Observatory | NSF's NOIRLab, Casilla 603, La Serena, Chile}
\email{atokovinin@ctio.noao.edu}

\begin{abstract}
Orbits  of inner  and outer  subsystems in  13 triple  or higher-order
stellar systems  are computed  or updated using  position measurements
and, in three cases, radial  velocities. The goal is to determine mutual
orbital  inclinations, period  ratios,  and masses  to complement  the
statistics of  hierarchical systems.  Effect  of the subsystems  on the
motion in the outer orbits  (wobble) is explicitly modeled to determine
inner mass  ratios.  Stars studied  here (HD 5408, 8036,  9770, 15089,
29310, 286955, 29316, 140538,  144362, 154621, 156034, 185655,  and 213235) are
bright and nearby (from 15\,pc  to 150\,pc). Their inner periods range
from 1.7 yr to  49 yr, and the outer periods from  83 to 2400 yr. Some
long-period  outer orbits are  poorly constrained.  Four astrometric
inner orbits and one outer orbit are computed for the first time.
\end{abstract} 

\keywords{binaries:visual; binaries:general}

\section{Introduction}
\label{sec:intro}

Architecture  of  hierarchical  stellar  systems  is  interesting  for
several  reasons.   Period   ratios,  mutual  orbit  orientation,  and
distribution of masses  are related to the processes  of formation and
early dynamical evolution of these systems, and therefore inform us on
the star  formation in general. Most  stars are born  in groups, hence
properties of binary and higher-order systems are very relevant.  Life
and death  of stars  is also affected  by their companions  in various
ways.   Close pairs  can  form within  hierarchical  systems by  their
dynamical  evolution  on  a  long  (compared to  the  formation)  time
scale.    Late   mergers    or   collisions    can    create   unusual
objects.  \citet{Hamers2020}  explored outcomes of dynamical  evolution  
using data on real hierarchical systems collected in the Multiple Star
Catalog \citep[MSC,][]{MSC}. 

This  study  is aimed  at  calculation  or  improvement of  visual  or
astrometric orbits of  the inner and outer subystems  belonging to the
same  hierarchy. There  are only  a few  dozen systems  where  this is
possible; the list  of 54 such objects with outer periods
less than  1000 yr is given  in \citet{moments}.  On the  one hand, most
outer systems with their typically long periods lack adequate coverage
needed for orbit  calculation. On the other hand,  resolution of inner
subsystems is possible  when they are relatively close  to the Sun and
their  periods are  not too  short.  Combination  of  both constraints
results in  a sample of  hierarchical systems with  comparable periods
and  separations and,  typically, with  similar masses. Obviously, the observed
sample is not  representative of  hierarchical systems  in the statistical
sense.   However,   hierarchies  with  comparable   periods  are  most
interesting  from  the   dynamical  perspective  because  interactions
between their subsystems might be important.

The observed motions in a  hierarchical system are modeled here by two
Keplerian orbits. In reality dynamical interactions cause evolution of
those orbits  over  time. These effects are too small  to be detected in
our systems. Given the orbital  elements and the estimated masses, the
long-term   dynamical   evolution    can   be   explored   numerically
\citep[e.g.][]{Xia2015,Hamers2020}, but this is  outside the scope of the present
study.

\begin{deluxetable*}{ccc cc lc lc l c }
\tabletypesize{\scriptsize}
\tablewidth{0pt}
\tablecaption{List of multiple systems
\label{tab:list} }
\tablehead{
\colhead{WDS} &
\colhead{HD} & 
\colhead{HIP} & 
\colhead{$V$} &
\colhead{$\varpi$\tablenotemark{a} } & 
\colhead{Outer\tablenotemark{b}} & 
\colhead{$P_{\rm out}$} &
\colhead{Inner\tablenotemark{b}} & 
\colhead{$P_{\rm in}$ } &
\colhead{Masses} &
\colhead{$N_{\rm comp}$}
 \\
\colhead{(J2000)} &        &  &
\colhead{(mag)} & 
\colhead{(mas)} & 
& \colhead{(yr)} &
& \colhead{(yr)} &
 \colhead{(\msun)}
}
\startdata
00568$+$6022 & 5408 & 4440    & 5.55  & 5.6 D  & V & 84.1  & V,A,S1 &  4.9 & (3.4+4.1)+3.4 & 4\\
01198$-$0031 & 8036 & 6226     & 5.87 & 7.76 G  & V  & 1000: & V,A & 27.1 & 2.5+(2.1+1.8)  & 3 \\
01350$-$2955 & 9770 & 7372     & 7.08 & 45.83 G & V  & 121  & V,A & 4.56 & (0.9+1.5)+0.6   & 4 \\
02291$+$6724 &  15089 & 11569   & 4.52 & 22.22 G & V & 2400: & V,A & 48.7 & (2.1+0.9)+1.3  & 5 \\ 
04375$+$1509 &  29310 & 21543    & 7.54 & 20.69 G & V,S1 & 125 & S1,A* & 2.0 & (1.1+0.3)+0.7 & 3\\  
04397$+$0952 & 286955 & 21710    & 9.19 & 34.04 G & V,S1 & 285 & S1,A* & 1.7 & (0.75+0.15)+0.5 & 4 \\
04400$+$5328 & 29316  & 21730    & 5.35 & 13.0 D & V    & 660: & V,A*  & 26.3   & (1.9+1.5)+1.5 & 3  \\
15440$+$0231  & 140538  & 77052   & 5.87 & 67.71 G & V & 900: & V,A & 6.6  & 1.0+(0.3+0.3) & 3 \\ 
16057$-$0617  & 144362  & 78849   & 6.36 & 12.03 G & V & 300: & V,A &  5.0 &  (1.6+1.6)+1.1& 4 \\
17066$+$0039  & 154621  & 83716   & 8.28  & 17.90 G & V  & 900: & V,A & 6.3 & 1.0+(0.7+0.7)& 3 \\
17157$-$0949  & 156034 & 84430    & 6.98  & 7.9  D  & V  & 137 & V & 5.3 &  1.7+(1.4+1.3) & 3\\
19453$-$6823  & 185655 & 97196    & 10.06 & 21.0 G  & V* & 84  & V,A & 4.4 & 0.8+(0.5+0.4)& 3 \\
22300$+$0426 & 213235 & 111062 &   5.51 & 18.84 G & V & 124 & V,A* & 2.1 & 1.7+(1.2+1.1)& 3 
\enddata
\tablenotetext{a}{Parallax codes: G --- Gaia \citep{Gaia}, D --- dynamical.}
\tablenotetext{b}{System types: V --- visual orbit, A --- astrometric
  orbit, S1 --- single-lined spectroscopic orbit, * --- first orbit. }
\end{deluxetable*}


Table~\ref{tab:list}  presents the  13 hierarchies  studied  here. Its
first three columns identify the systems by the Washington Double Star
\citep[WDS,][]{WDS}  codes  based  on  J2000 positions,  HD,  and  HIP
numbers.   The  combined  visual  magnitudes  come  from  Simbad,  the
parallaxes are either measured  by Gaia \citep{Gaia} or estimated from
the masses and orbits as explained below. Types of the orbits (visual,
astrometric,  or spectroscopic)  and periods  of the  outer  and inner
subsystems are  given. Astrometric or  visual orbits computed  for the
first time  here are marked  by asterisks.  Some of  these hierarchies
have more than three components, but they are effectively treated here
as triples.  Components' masses  of these triples   and  the total
  number of components  are given in the last two columns (inner subsystem
  in brackets).

Most   of   these  bright   and   nearby   hierarchies  had   previous
determinations of their inner and  outer visual orbits or even a joint
analysis  \citep[e.g.][]{Hei1996b},  so why  a  new  study is  needed?
First,  outer and  inner  orbits are  fitted  here jointly,  including
the wobble caused by the subsystem.   Second, recent speckle data are used
and  appropriate  weights are  applied;  radial  velocities (RVs)  are
included  when  available to  improve  the  accuracy.   Masses of  the
components  are  estimated  using  new Gaia  parallaxes  and  absolute
magnitudes  (or isochrones  for evolved  stars).   Consistency between
parallaxes, orbits,  and estimated mass sums is  achieved; elements of
poorly  constrained long-period orbits  are chosen  to match  the mass
sums.  Finally, accuracy of the orbital elements is evaluated, as well
as the  reliability of mutual  inclination and period  ratios computed
from  these  orbits. Two  astrometric  inner  orbits of  spectroscopic
subsystems, two visual inner orbits, and one outer orbit are first-time
solutions.   This  study continues  previous  work  on  the orbits  of
hierarchical systems \citep{TL2017,TL2020,twins}.

Components of multiple systems are designated by sequences of letters
and numbers, as in the WDS. Subsystems are denoted as two components
joined by a comma. Thus, A,B means a subsystem with components A and B,
while AB stands for the composite component (center of mass) in a
wider pair AB,C. I try to avoid using the outdated discoverer codes
given in the WDS, but mention them for completeness in the notes on 
individual systems. 

Section~\ref{sec:data} presents the methods briefly. The main part of
the paper is Section~\ref{sec:sys}, where each hierarchy is discussed
individually,  and masses and other parameters are estimated. A short summary
in Section~\ref{sec:sum} closes the paper.  

\section{Data and methods}
\label{sec:data}

The orbit of  a visual  binary is defined  by the seven  Campbell elements
$P,T,e,a,  \Omega, \omega,  i$ (in  standard notation).  Fitting these
elements  to  position measurements  $\theta(t)$  and  $\rho(t)$ is  a
classical problem  with many  solutions given in the  literature. I  use the
least-squares fit with weights inversely proportional to the square of
measurement errors  $\sigma^2_i$. However,  the actual errors  are not
Gaussian  and are not  well known,  leaving room  for  subjective choices;
different orbits can  be derived from the same  data, depending on the
method and weights.

The observed motion  in a resolved triple system  (e.g. Aa,Ab and A,B)
is modeled  here by  two Keplerian orbits  and the wobble  factor $f$,
hence there are  15 free parameters; more parameters  (e.g.  the inner
and  outer  RV  amplitudes  $K_1$,  $K_3$,  and  the  system  velocity
$\gamma$)  are fitted  when the  RV measurements  are  available.  The
angles $\Omega$ and  $\omega$ are chosen to fit the  RV of the primary
component, otherwise  both can be changed  simultaneously by 180\degr.
The wobble  factor is  the ratio  of the wave  amplitude in  the outer
orbit caused by  the subsystem to the inner  semimajor axis.  When the
outer positions  of Aa,B refer to  the resolved inner  subsystem, $f =
q/(1+q)$ is directly related to  the inner mass ratio $q_{\rm Aa,Ab}$;
it is  also called  the fractional mass  \citep{Hei1996b,TL2017}. 
  When the positions of the outer component B are measured relative to
  the photo-center of the unresolved inner subsystem, e.g. A,B, then 
the wobble factor $f^* = f -  r/(1+r)$ depends on the light ratio $r =
10^{-0.4 \Delta m}$.  For example,  if the inner subsystem consists of
two identical stars, $f = 0.5$ and $f^* = 0$. When the inner subsystem
is not resolved, I set its semimajor axis to the value calculated from
the period and mass and fit the wobble factor to determine the axis of
the astrometric orbit.

I    use   the    IDL    code   {\tt    orbit3.pro}\footnote{Codebase:
  \url{http://dx.doi.org/10.5281/zenodo.321854}}  to  fit both  orbits
simultaneously  \citep{TL2017,twins,TL2020}.   Its  modification  {\tt
  orbit4.pro} distinguishes between  outer positions measured from
  a resolved and unresolved inner  system;  the factor $f$ is fitted,
while the ratio $f^*/f$ is fixed to a value deduced from a model, e.g.
zero for  equal-mass subsystems.  Often  accurate speckle measurements
refer to  the resolved subsystem Aa,B, while  the remaining unresolved
micrometer measurements  are so crude that the  photo-center wobble is
lost  in their  noise, making  the distinction  between $f$  and $f^*$
irrelevant.

The  position  measurements  were  retrieved  from  the  WDS  database
\citep{WDS}  on my request  and, where  possible, complemented  by the
recent speckle  interferometry at the  Southern Astrophysical Research
(SOAR)  4.1 m  telescope \citep[][and  references  therein]{SOAR}. The
latest SOAR  observations are made  in 2020.8.  The  SOAR measurements
and the  data from  telescopes with apertures  of 6\,m and  larger are
assigned errors of 1--2 mas, speckle interferometry at other 4 m class
telescopes is assumed to have  errors of 5 mas. For smaller apertures,
the  errors   are  correspondingly  larger.    The  visual  micrometer
measurements are  assumed to have errors from  0\farcs05 to 0\farcs25,
depending on the separation.  These errors are poorly known, while the
old  data   vary  in  quality, depending  on  the observer.   I
liberally  deleted the largest  historical outliers  and down-weighted
some other  micrometer measurements.  The system of  weights used here
privileges the speckle data and  is different from the weights adopted
by the orbit catalog \citep{VB6}.  The existing outer orbits of triple
systems were typically computed  by simple Keplerian fits, neglecting
the  wobble; its explicit modeling here improves the accuracy.

Some long-period  outer orbits considered  here have only a  short arc
covered  by the  observations,  resulting in  the loosely  constrained
elements.   Previously computed  orbits of  these pairs  have  grade 5
(preliminary) in the catalog \citep{VB6}.  While the standard approach
uses  only  the  position  measurements, I  consider  here  additional
information on  the mass  sum, parallax, and,  in some cases,  the RVs.
Large errors  of some  elements (e.g.  period  or inclination)  in the
initial, unconstrained,  fit indicate that  a wide range of  orbits is
compatible  with the  data. Poorly  defined elements  usually strongly
correlate with other elements.  I  fix some poorly defined elements to
satisfy  the  additional constraints  on  the  mass  sum and  fit  the
remaining elements; the formal errors of the fitted elements in such orbits
are in fact  only lower limits.  Even uncertain  long-period orbits of
hierarchical systems are useful  because they accurately represent the
observations, serve  as a reference for measuring the  wobble, and allow
calculation  of the mutual  inclination $\Phi$.   In most  cases,  the true
ascending nodes of  the inner and/or outer orbits  are not known, and
two alternative values of $\Phi$ are possible \citep{moments}.

Masses of main-sequence stars are estimated from their absolute visual
magnitudes using standard relations \citep{Pecaut2013}. The parallaxes
are measured by Gaia \citep{Gaia}. However, its current data release
does not account for the non-linear motion of binaries. Gaia parallaxes
and proper motions (PMs) of binary stars may have large errors and
biases, and the relation between these effects and the orbital parameters is
complex \citep{Penoyre2020}. The short-term PM of
binaries measured by Gaia often differs from their long-term PM deduced from
the Hipparcos and Gaia positions. This PM anomaly \citep{Brandt2018}
is a manifestation of astrometric acceleration; sometimes it helps
to constrain or verify the orbits. I use the more accurate Gaia parallaxes
of other (non-binary) components of multiple systems when they are
available. Otherwise, dynamical parallaxes computed from the orbits and
the estimated masses are  currently the best distance measurements.

All orbit plots in  the next section (e.g. Figure~\ref{fig:00568}) are
similar.  They show the outer orbit, including the wobble, by the wavy
line.  Measurements of the  outer pair (asterisks if resolved, squares
and crosses otherwise) are connected  to the expected positions on the
orbit  by  short  dotted  lines.  The  scale  is  in  arcseconds,  the
orientation  is  standard  (north  up,  east left),  and  the  primary
component (large  asterisk) is located at the  coordinate origin.
  The orbit  of the inner subsystem  is plotted at the  center, on the
  same scale,   by the magenta  dash ellipse and  triangles.  Even if
the inner subsystem  belongs to the secondary component,  its orbit is
plotted  around  the  center;  in  this case,  the  wobble  factor  is
negative.  Each  plot also contains a  graph representing  the hierarchical
structure of the multiple system (the mobile diagram).  Masses of stars
and orbital periods in italics accompany these diagrams.

\begin{deluxetable*}{ l  c rrr rrr r r r }
\tabletypesize{\scriptsize}
\tablewidth{0pt}
\tablecaption{Orbital Elements \label{tab:orb}}
\tablehead{
\colhead{WDS} &
\colhead{System} &
\colhead{$P$} & 
\colhead{$T  $} &
\colhead{$e$} & 
\colhead{$a$} & 
\colhead{$\Omega$} &
\colhead{$\omega$} &
\colhead{$i$}  &
\colhead{$f$} &
\colhead{$K$} 
 \\
 & &   
\colhead{(yr)} & 
\colhead{(yr)} &
\colhead{ } & 
\colhead{($''$)} & 
\colhead{(deg)} &
\colhead{(deg)} &
\colhead{(deg)} & &
\colhead{(\kms)}
}
\startdata
00568$+$6022 & Aab,Ac & 4.899 & 2003.565 & 0.260   & 0.0309 & 146.8 & 279.2        & 47.3 &    0.346 & 11.18 \\
             &   & $\pm$0.009 &$\pm$0.038&$\pm$0.017&$\pm$0.0012&$\pm$2.5 &$\pm$2.9&$\pm$3.2 &$\pm$0.065 &$\pm$0.24 \\
00568$+$6022 & A,B & 84.10 & 1953.59 & 0.225 & 0.237 & 171.5 & 343.5 & 53.6 & \ldots & 4.30  \\
             &     & $\pm$0.84 & $\pm$0.83 & $\pm$0.010 & $\pm$0.006 & $\pm$0.9 & $\pm$3.6 & $\pm$0.9 & \ldots& fixed \\
01198$-$0031 & B,C & 27.13 & 1994.92 & 0.205 & 0.1108 & 41.2 & 343.2 & 18.5 &   $-$0.466& \ldots\\
             &     & $\pm$0.08 & $\pm$0.19 & $\pm$0.010 & $\pm$0.0014 & $\pm$11.4 & $\pm$13.0 & $\pm$3.5 & $\pm$0.015& \ldots\\
01198$-$0031 & A,BC & 1000 & 1686.7 & 0.540 & 1.450 & 123.1 & 102.8 & 37.9 & \ldots & \ldots\\
             &     & fixed & $\pm$17.4 & fixed & $\pm$0.021 & $\pm$9.8 & $\pm$5.5 & $\pm$1.1 & \ldots& \ldots \\
01350$-$2955 & A,B & 4.5595 & 2010.127 & 0.3142 & 0.1669 & 168.6 & 298.9 & 11.9 &    0.640& 2.92 \\
             &     & $\pm$0.0014 & $\pm$0.006 & $\pm$0.0022 & $\pm$0.0006 & $\pm$6.1 & $\pm$6.0 & $\pm$1.7 & $\pm$0.012& $\pm$0.69 \\
01350$-$2955 & AB,C & 120.88 & 1955.24 & 0.230 & 1.571 & 132.5 & 54.3& 36.5 & \ldots & \ldots\\
             &     & $\pm$0.83 & $\pm$0.95 & $\pm$0.011 & $\pm$0.018 & $\pm$3.0 & $\pm$4.0 & $\pm$1.2 & \ldots& \ldots \\
02291$+$6724 & Aa,Ab & 48.72 & 1993.21 & 0.637 & 0.423 & 176.6 & 328.2 & 148.2 &    0.322& \ldots\\
             &     & $\pm$0.45 & $\pm$0.05 & $\pm$0.004 & $\pm$0.004 & $\pm$1.8 & $\pm$1.9 & $\pm$1.3 & $\pm$0.013& \ldots\\
02291$+$6724 & A,B & 2400 & 940 & 0.40 & 6.50 & 188.0 & 113.3 & 102.9 & \ldots& \ldots \\
             &     & fixed & $\pm$47 & fixed & fixed & $\pm$0.9 & $\pm$3.4 & $\pm$0.3 & \ldots& \ldots \\
04375$+$1509 & Aa,Ab & 2.0097 & 1999.818 & 0.405 & 0.037 & 319.3 & 339.2 & 144.0 &    0.223& 3.85 \\
             &     & $\pm$0.0009 & $\pm$0.011 & $\pm$0.014 & fixed & $\pm$11.1 & $\pm$3.0 & fixed & $\pm$0.043& $\pm$0.06 \\
04375$+$1509 & A,B & 124.9 & 2049.6 & 0.20 & 0.667 & 324.0 & 260.5 & 73.3 & \ldots & 2.23  \\
             &     & $\pm$22.6 & $\pm$12.5 & fixed & $\pm$0.080 & $\pm$2.2 & $\pm$44.3 & $\pm$1,8 & \ldots& $\pm$0.23 \\
04397$+$0952 & Aa1,Aa2 & 1.6732 & 2013.958 & 0.426 & 0.046 & 286.5 & 267.7 & 62.7 &    0.170& 4.47 \\
             &     & $\pm$0.0005 & $\pm$0.013 & $\pm$0.018 & fixed & $\pm$10.2 & $\pm$3.6 & $\pm$14.3 & $\pm$0.030& $\pm$0.10\\
04397$+$0952 & Aa,Ab & 285 & 1999.8 & 0.19 & 1.637 & 335.7 & 253.0 & 78.6 & \ldots& 1.91  \\
             &     & fixed & $\pm$76 & $\pm$0.09 & $\pm$0.136 & $\pm$4.9 & $\pm$145 & $\pm$1.5 & \ldots & $\pm$0.57 \\
04400$+$5328 & A,B & 26.34 & 1988.98 & 0.846 & 0.1727 & 12.6 & 42.9 & 138.4 &    0.446& \ldots\\
             &     & $\pm$0.05 & $\pm$0.03 & $\pm$0.005 & $\pm$0.0023 & $\pm$2.5 & $\pm$2.6 & $\pm$1.2 & $\pm$0.009& \ldots\\
04400$+$5328 & AB,C & 660 & 2011.7 & 0.405 & 1.666 & 286.2 & 105.1 & 132.5 & \ldots& \ldots \\
             &     & fixed & $\pm$2.7 & $\pm$0.015 & $\pm$0.019 & $\pm$1.8 & $\pm$5.4 & $\pm$1.9 & \ldots& \ldots \\
15440$+$0231 & Ba,Bb & 6.57 & 2020.09 & 0.357 & 0.189 & 21.4 & 230.0 & 70.4 &   $-$0.500& \ldots\\
             &     & $\pm$0.29 & $\pm$0.08 & $\pm$0.038 & $\pm$0.008 & $\pm$2.7 & $\pm$9.3 & fixed & fixed& \ldots\\
15440$+$0231 & A,B & 900 & 1936.1 & 0.435 & 7.20 & 54.9 & 330.9 & 138.1 & \ldots& \ldots \\
             &     & fixed & $\pm$9.9 & $\pm$0.030 & fixed & $\pm$7.4 & $\pm$3.4 & $\pm$1.5 & \ldots& \ldots \\
16057$-$0617 & Aa,Ab & 4.999 & 2013.16 & 0.370 & 0.0521 & 117.3 & 8.7 & 127.0 &    0.485& \ldots\\
             &     & $\pm$0.005 & $\pm$0.06 & $\pm$0.009 & $\pm$0.0011 & $\pm$2.5 & $\pm$5.7 & $\pm$1.7 & $\pm$0.053& \ldots\\
16057$-$0617 & A,B & 300 & 2048.5 & 0.611 & 0.877 & 107.6 & 141.0 & 155.0 & \ldots& \ldots \\
             &     & fixed & $\pm$4.1 & $\pm$0.033 & $\pm$0.009 & $\pm$24.0 & $\pm$26.3 & fixed & \ldots& \ldots \\
17066$+$0039 & Ba,Bb & 6.311 & 2013.04 & 0.400 & 0.0700 & 2.4 & 195.2 & 29.9 &   $-$0.521& \ldots\\
             &     & $\pm$0.031 & $\pm$0.03 & $\pm$0.010 & $\pm$0.0014 & $\pm$6.2 & $\pm$7.0 & $\pm$2.9 & $\pm$0.031& \ldots\\
17066$+$0039 & A,B & 900 & 1963.04 & 0.596 & 2.264 & 97.7 & 356.9 & 40.0 & \ldots& \ldots \\
             &     & fixed & $\pm$1.36 & $\pm$0.002 & $\pm$0.023 & $\pm$3.2 & $\pm$1.8 & fixed & \ldots& \ldots \\
17157$-$0949 & Ba,Bb & 5.257 & 2011.08 & 0.46 & 0.033 & 52.3 & 69.0 & 138.6 &   $-$0.436& \ldots\\
             &     & $\pm$0.052 & $\pm$0.59 & $\pm$0.11 & $\pm$0.008 & $\pm$11.7 & $\pm$22.5 & $\pm$24.9 & $\pm$0.040& \ldots\\
17157$-$0949 & A,B & 137.1 & 2016.09 & 0.357 & 0.344 & 26.5 & 280.5 & 141.4 & \ldots& \ldots \\
             &     & $\pm$1.8 & $\pm$0.26 & $\pm$0.007 & $\pm$0.002 & $\pm$1.2 & $\pm$1.7 & $\pm$1.1 & \ldots& \ldots \\
19453$-$6823 & Ba,Bb & 4.40 & 2017.22 & 0.850 & 0.0544 & 148.2 & 218.9 & 127.4 &   $-$0.344& \ldots\\
             &     & $\pm$0.09 & $\pm$0.06 & fixed & $\pm$0.0052 & $\pm$6.4 & $\pm$10.3 & $\pm$7.6 & $\pm$0.032& \ldots\\
19453$-$6823 & A,B & 84.4 & 2001.55 & 0.252 & 0.475 & 204.0 & 336.2 & 140.0 & \ldots& \ldots \\
             &     & $\pm$1.8 & $\pm$0.86 & $\pm$0.016 & $\pm$0.027 & $\pm$2.6 & $\pm$1.9 & fixed & \ldots& \ldots \\
22300$+$0426 & Ba,Bb & 2.113 & 2016.60 & 0.193 & 0.0412 &266.9 & 151.3 & 113.1 &   $-$0.456& \ldots\\
             &     & $\pm$0.017 & $\pm$0.06 & $\pm$0.027 & $\pm$0.0012 & $\pm$2.0 & $\pm$9.0 & $\pm$1.7 & $\pm$0.021\\
22300$+$0426 & A,B & 123.53 & 2035.38 & 0.515 & 0.714 & 117.1 & 214.6 & 89.94 & \ldots & \ldots\\
             &     & $\pm$1.26 & $\pm$0.92 &$\pm$0.016  & $\pm$0.009 & $\pm$0.1 & $\pm$1.5 & $\pm$0.04 & \ldots & \ldots
\enddata
\end{deluxetable*}

Elements  of visual  orbits (including  the wobble  factor)  and their
errors are listed in Table~\ref{tab:orb}. The last column gives the RV
amplitudes when the  RVs were used; the systemic  velocities are given
in the  discussion of individual  systems.  As only published RVs are
used, they are not  duplicated here.  The position measurements, their
adopted  errors,  and  residuals   to  the  orbits  are  assembled  in
Table~\ref{tab:speckle}, published in full electronically.

\begin{deluxetable*}{r l l rrr rr l}    
\tabletypesize{\scriptsize}     
\tablecaption{Position measurements and residuals (fragment)
\label{tab:speckle}          }
\tablewidth{0pt}                                   
\tablehead{                                                                     
\colhead{WDS} & 
\colhead{Syst.} & 
\colhead{Date} & 
\colhead{$\theta$} & 
\colhead{$\rho$} & 
\colhead{$\sigma_\rho$} & 
\colhead{(O$-$C)$_\theta$ } & 
\colhead{(O$-$C)$_\rho$ } &
\colhead{Ref.\tablenotemark{a}} \\
 & & 
\colhead{(JY)} &
\colhead{(deg)} &
\colhead{(\arcsec)} &
\colhead{(\arcsec)} &
\colhead{(deg)} &
\colhead{(\arcsec)} &
}
\startdata
00568+6022 & Aa,Ab &  1994.7183 &    161.9 &   0.0302 &   0.0020 &      1.8 &   0.0009 & s \\
00568+6022 & Aa,Ab &  1998.7731 &     98.3 &   0.0200 &   0.0020 &      9.7 &   0.0032 & s \\
00568+6022 & Aa,Ab &  1998.7760 &     69.8 &   0.0160 &   0.0020 &    $-$19.2 &  $-$0.0009 & s \\
00568+6022 & Aa,Ab &  2000.8744 &    228.1 &   0.0250 &   0.0020 &     $-$6.2 &  $-$0.0013 & s \\
00568+6022 & Aa,Ab &  2000.8770 &    228.9 &   0.0250 &   0.0020 &     $-$5.5 &  $-$0.0013 & s \\
00568+6022 & Aa,B &  1994.7183 &    339.7 &   0.2720 &   0.0050 &     $-$0.2 &   0.0011 &  s\\
00568+6022 & Aa,B &  1994.7183 &    339.7 &   0.2725 &   0.0050 &     $-$0.2 &   0.0016 &  s 
\enddata 
\tablenotetext{a}{
A: adaptive optics;
C: CCD measurement; 
G: Gaia;
H: Hipparcos;
J: Finsen's eyeiece micrometer;
M: visual micrometer measurement;
P: photographic measurement;
S: speckle interferometry at SOAR;
s: speckle interferometry at other telescopes.
}
\end{deluxetable*}

\section{Individual Systems}
\label{sec:sys}

\subsection{00568$+$6022 (ADS 784, HR 266)}

\begin{figure}
\epsscale{1.0}
\plotone{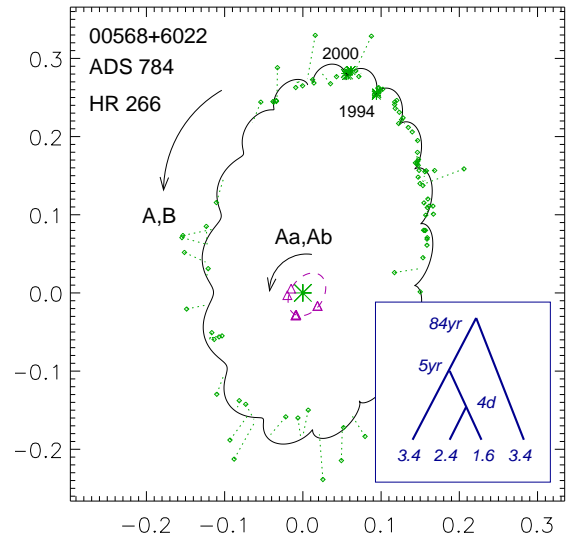}
\caption{Outer and intermediate orbits of ADS 784 (periods 84 yr and 5
  yr). In this and following figures, the wavy line marks the outer
  orbit. Measured positions (asterisks and crosses) are connected to
  the expected positions by short dotted lines. Scale in arcseconds,
  north up, east left, primary component at coordinate origin. The
  inner orbit is plotted at the center on the same scale (magenta
  ellipse and triangles). The blue graph shows the hierarchical structure of
  the complete system, with approximate periods and masses in italics. 
\label{fig:00568}  }
\end{figure}

HR~266 (ADS 784) is an  early-type quadruple system. The outer pair is
a  84 yr visual  binary BU~1099.  Its orbit,  fully covered  and quite
accurate, is  only slightly updated  here.  \citet{Cole1992} published
a detailed study  of this system, including spectroscopic  orbits of the
inner 4.2 day double-lined spectroscopic  binary and its 5 yr orbit 
   in  an  intermediate  subsystem,   which  was also  discovered
independently by the wavy motion of  A,B. Therefore, this is a
3+1 hierarchy where  a close binary has a  tertiary companion and this
triple    is   orbited    by   another,    more    distant visual  companion
(Figure~\ref{fig:00568}).

\citet{Schoeller1998} reported  direct  resolution of  the intermediate
pair Aa,Ab  at 25  mas separation, near  the diffraction limit  of the
6\,m telescope.  While Cole et al. attributed the spectroscopic triple
to the secondary visual component B, in  fact the component A has been resolved.
\citet{Balega1999}  revise the  system  model where  the inner  triple
belongs to  the component A and  show that it does  not contradict the
spectroscopy of Cole et al.  However, their model identifies the inner
spectroscopic  binary  with  Aa   and  predicts  zero  wobble  in  the
unresolved measurements  of A,B because  the mass and light  ratios of
the  5  yr subsystem  are  nearly  equal  (about 0.5),  canceling  the
photo-center wobble.  I repeated their  analysis and reached  the same
conclusion  that   contradicts  the  observed  wavy   motion  of  A,B.
\citet{Doc2006}  computed  a crude  visual  orbit  of  the 5  yr  pair
(without  using  the RVs)  and  also  estimated  masses from the absolute
magnitudes. They  found that most of  the mass is  concentrated in the
inner triple, but still call this component B.

Here  I attempt  to clarify  the  confusion regarding  this system  by
attributing the  inner 4-day pair to  the secondary component  Ab and
denoting it as Ab1,Ab2.  Speckle observations published since the last
analysis by Balega et al. and the RVs of the center-of-mass of Ab from
Cole et al. are used to determine the combined orbit of Aa,Ab jointly
with the outer orbit of A,B. 

The  only  resolved  measurements of  Aa,Ab  were  taken  at the  6  m
telescope on three  epochs, 1994.7, 1998.8, and 2000.9  (I swapped the
published quadrant in 2000.9). Micrometric and speckle measurements of
the outer pair A,B (without resolving the subsystem) from 1889 to 2010
cover  131 yr  or  1.56  outer periods.   The  photo-center wobble  is
assumed to be the $f^*/f =0.32$ fraction of the full wobble to fit the
wave  in the unresolved  measurements.  The  first speckle-astrometric
orbit of  the 5-yr pair computed by  Cole et  al. was based  only on the
unresolved measurements  made before 1990 and had  a  wobble amplitude
of 4\,mas.

Differential  speckle photometry by  \citet{Schoeller1998} in  the $V$
band indicates that the magnitudes of Aa and B are similar within 0.1 mag,
while Ab is 0.7 mag fainter. The inferred magnitude difference between
A and B, 0.45 mag, agrees with its multiple measurements. So, Aa and B
have  similar  luminosities and  masses  and,  together, dominate  the
combined spectrum. This explains why  Cole et al. could not detect the
lines of the  5 yr secondary and why they  attributed the 4-day binary
to B.  Using the 50  Myr PARSEC isochrone \citep{PARSEC}, I found that
the magnitude  difference between Ab1 and  Ab2 of 1.4  mag matches the
spectroscopic  mass   ratio  $q_{\rm  Ab1,Ab2}  =   0.67$.  With  this
assumption,  the  visual  magnitudes  of all  components  are  defined
(Table~\ref{tab:ADS784}). Furthermore, I  adopt the dynamical parallax
of 5.6\,mas  derived from the A,B  orbit and the expected  mass sum of
$\sim$11  \msun.  The  corresponding  spectral types,  also listed  in
Table~\ref{tab:ADS784}, agree with the spectral classification by Cole
et al., except the A7V component Ab2 which they classified as A1V. The
two most massive members of this  system, B and Aa, have spectral type
B7V and are fast rotators.

\begin{deluxetable}{l c c cc  }
\tabletypesize{\scriptsize}     
\tablecaption{Parameters of Components of ADS~784
\label{tab:ADS784} }  
\tablewidth{0pt}                                   
\tablehead{                                                                     
\colhead{Parameter} & 
\colhead{Ab1} &
\colhead{Ab2} &
\colhead{Aa} &
\colhead{B} 
}
\startdata
$V$ (mag)                    &     7.53  &  8.83 &  6.57 &  6.57 \\
${\cal M}$ (${\cal M}_\odot$) &     2.44  &  1.64 &  3.39 &  3.39 \\
Spectral type               &      B9V    &   A7V  &   B7V  &    B7V 
\enddata
\end{deluxetable}

The adopted parallax of 5.6\,mas leads to the inner mass sum of 7.1
\msun, in reasonable agreement with 7.5 \msun estimated from the
isochrone. The RV amplitude of Ab and its mass, 4.1 \msun, correspond
to the 3.1 \msun mass for Aa that matches the isochrone mass
approximately. Cole et al. noted the large mass of the unseen
secondary in the 5-yr subsystem and suggested that it could be a pair
of low-mass stars. This hypothesis is no longer viable because speckle
interferometry resolved the subsystem Aa,Ab. 

The system  model in Table~\ref{tab:ADS784} implies the  mass ratio in
the  5-yr subsystem  $q_{\rm Aa,Ab}  = 1.2$  which, together  with the
measured  light  ratio $r=0.5$,  results  in  the photo-center  wobble
factor $f^* =  0.20$ and the wobble amplitude  of 6.2\,mas, similar to
the amplitude estimated by Cole  et al. However, the calculated wobble
factor for  the resolved Aa,B positions, $f=0.55$,  disagrees with the
measured $f  = 0.35 \pm 0.07$.  I suspect that the  measurement of $f$
based on only three epochs is questionable.

The  outer  orbit  and  the  estimated masses  correspond  to  the  RV
amplitudes in the outer orbit of $K_3= 4.3$ and $K_4=9.5$ \kms.  No RV
trend associated with  the outer orbit is obvious  in the RVs measured
by Cole et al. Adopting $K_3=4.3$ \kms, I obtain the systemic velocity
of  $-7.6$ \kms  or $-9.5$  \kms, depending  on the  node of  A,B. The
choice  of the  outer node  also defines  the mutual  inclination. The
model  proposed  here attributes  the  4-day  subsystem  to Ab,  hence
$\omega_{\rm Ab} = 279\fdg5$ refers  to the secondary component of the
5-yr  subsystem. If  $\omega_B =  327\fdg9$  in the  outer orbit  also
refers to the secondary, the  mutual inclination between the orbits of A,B
and Aa,Ab  is $\Phi  = 20\degr \pm  2\degr$, otherwise the  orbits are
nearly perpendicular,  $\Phi = 90\degr$. Continued  RV monitoring will
define the sign of the long-term  RV trend and hence the correct node.
The inclination of the 4-day subsystem is $i_{\rm Ab1,Ab2} = 63\degr$.

Relation between  ADS 784 and $\gamma$~Cas  (WDS J00567+6043, HD~5394,
HR 264), the  archetype of Be stars, was  noted by \citet{Mamajek2017}.
The Hipparcos parallax of $\gamma$~Cas is 5.32$\pm$0.56 mas and its RV
is  $-7.4$ \kms;  the  PMs of  these  systems also  match.  Note  that
$\gamma$~Cas   is  a  triple   system  consisting   of  the   203  day
spectroscopic  pair,  a  60  yr  astrometric subsystem,  and  a  faint
physical  companion  at  2\farcs05.   The projected  distance  between
HR~266 and  HR~264 is  about 1\,pc,  too large for  a bound  pair (its
orbital period  would be $\sim$20\,Myr). More likely,  these young and
massive hierarchical systems are linked together by  their common origin.

\subsection{01198$-$0031 (42 Cet, ADS 1081)}

\begin{figure}
\epsscale{1.0}
\plotone{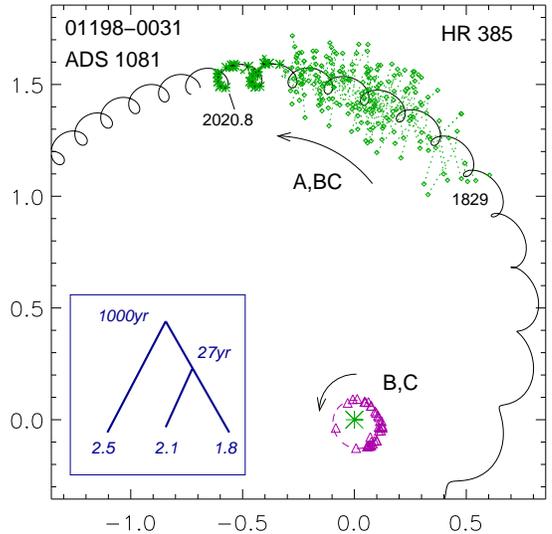}
\caption{Orbits of 42~Cet (ADS 1081), periods 1000 yr and 27 yr. 
\label{fig:01198} }
\end{figure}

The bright star  HR~385 (42~Cet, ADS~1081) consists of an evolved G8III
primary and the tight pair of  nearly equal A7V stars B and C (FIN~337
BC). The inner 27 yr orbit is very well defined \citep{Msn2010c}, unlike the
long-period outer  orbit of STF~113  A,BC that lacks  coverage despite
measurements  available since  1829.  Gaia  gives rather  accurate and
consistent parallaxes of stars A  and BC, and their average, 7.76\,mas,
defines the distance to the system. 

The differential  photometry at SOAR  gives $\Delta y_{\rm AB}  = 0.96
\pm 0.15$ mag, hence  the $V$ magnitudes of B and C  are 7.50 and 8.46
mag.  The  orbit of  B,C gives a  mass sum  of 3.96 \msun.  The wobble
factor $f  =-0.466 \pm 0.016$ implies  $q_{\rm BC} =  0.87$, hence the
measured masses  of B and C  are 2.11 and 1.84  \msun, somewhat larger
than 1.81 and 1.46 \msun  estimated from the absolute mangitudes.

The  latest outer  orbit with  $P=650$\,yr \citep{Zir2015a}  yields an
unrealistically large outer mass sum  of 11.8 \msun vs.  6.5 \msun
  estimated  from the  absolute  magnitudes. The  new  outer orbit  in
Figure~\ref{fig:01198} has a period  arbitrarily fixed at 1000\,yr and
the fixed eccentricity chosen to  obtain the target mass sum. Historic
observations that deviated  by more than 0\farcs2 from  the orbit were
removed from  the final  fit. A free  fit has huge  errors, indicating
that the data do not  really constrain the outer orbit.  The tentative
outer orbit serves mostly as a reference to measure the wobble factor,
and the formal errors of its non-fixed elements are only lower limits.
The two  mutual inclinations are  almost equal, about  40\degr.

\subsection{01350$-$2955 (GJ 60)}

\begin{figure}
\epsscale{1.0}
\plotone{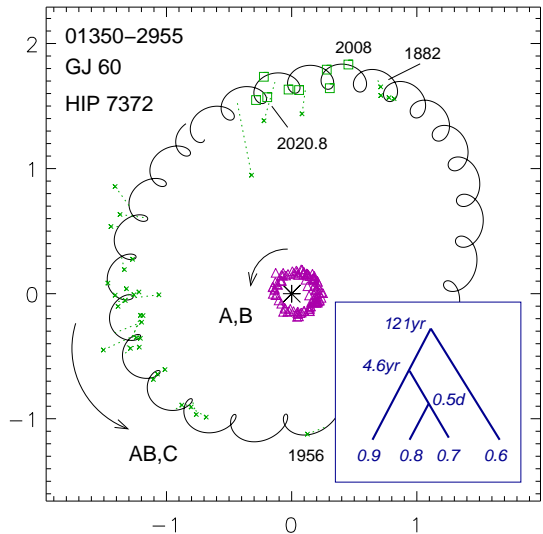}
\caption{Orbits of GJ 60 AB,C (BU 1000, 121 yr) and A,B  (DAW 31, 4.56 yr).
\label{fig:01350}  }
\end{figure}

 The well-studied, solar  neighborhood quadruple system GJ~60 consists
 of     four     K-type      dwarfs     with     comparable     masses
 (Figure~\ref{fig:01350}).    The    Gaia   parallax   of    star   C,
 45.83$\pm$0.18 mas,  defines the  system's distance (the  parallax of
 AB, 38.4\,mas,  is strongly biased).   Star B is an  eclipsing binary
 (designated  BB~Scl)  with a  period  of  0.4765  day.  This  bright,
 nearby,  and  chromospherically  active  star  is  mentioned  in  149
 references.

This is a rare case where both  the outer 121 yr and the inner 4.56 yr
visual orbits are well constrained and of good quality.  New, slightly
updated,  orbits are  obtained by  the  full unconstrained  fit of  15
elements. Accurate SOAR data  from 2008--2020 define the wobble factor
$f = 0.640 \pm 0.012$, which clearly shows that B is more massive than
A, $q_{\rm A,B} = 1.78$.  The  inner mass sum is 2.33 \msun, hence the
measured masses of A and B are 0.84 and 1.49 \msun, respectively.  The
chromospherically active eclipsing  subsystem Ba,Bb consists of almost
equal   mass stars with  $q_{\rm Ba,Bb}=0.97$ \citep{Watson2001}.
The estimated mass of C, 0.56 \msun, leads to the total system mass of
2.89 \msun, which  agrees well with 2.75 \msun  deduced from the outer
orbit and the parallax.  If the  eclipsing nature of B were not known,
the  subsystem  Ba,Bb  could  be  independently  discovered  from  the
excessive mass of B.

The period ratio  of the visual pairs is  26.5$\pm$0.2.  The crude RVs
of component A plotted in Figure~5 of \citet{Watson2001} indicate that
the node of the 4.6 yr  orbit listed here corresponds to the component
A.  The true  node of the outer 121 yr orbit  remains unknown, and the
small  estimated  RV  amplitude   of  1.3  \kms  makes  such  prospect
unlikely. This  leaves two options for the  mutual inclination, either
28\degr$\pm$2\degr ~or 47\degr. The  first choice appears more likely,
considering the moderate inner  and outer eccentricities (0.31 and 0.23).
This  hierarchical  system  resembles  other planetary  type  low-mass
hierarchies   \citep{twins},  but  its   inner  eclipsing   system  is
definitely  not coplanar  with  the low-inclination  middle and  outer
orbits. Therefore, the eclipsing subsystem BB~Scl should be precessing
and this might  be detectable by the variable  eclipse depth. Analysis
of the eclipse time variation  can bring additional information on the
orbits, but it is outside the scope of this study.

\subsection{02291$+$6724 ($\iota$ Cas, ADS 1860)}

\begin{figure}
\epsscale{1.0}
\plotone{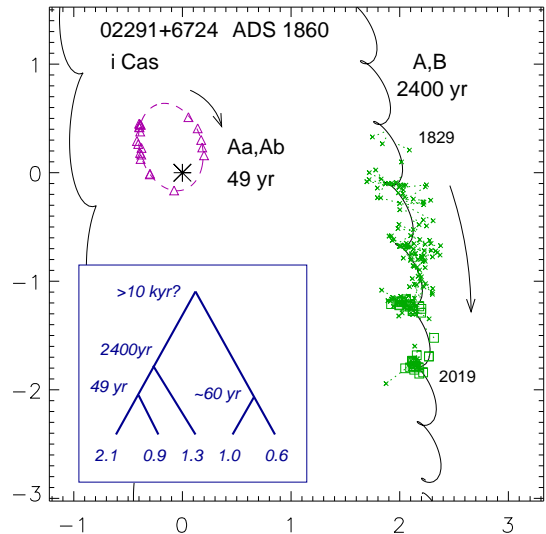}
\caption{Orbits of $\iota$ Cas (ADS 1860). A fragment of the outer (2400 yr) orbit and the inner (49 yr) orbit.
\label{fig:02291}  }
\end{figure}

$\iota$~Cas  (HR~707,  ADS  1860)  is a  well-known  quintuple  system
(Figure~\ref{fig:02291}) with  an A5Vp magnetic  primary.  Three stars
A, B, and C were resolved by W.~Struve in 1828 (STF 262) at comparable
separations and earlier by W.~Hershel (his inaccurate measurements are
not used  here). The  47 yr astrometric  subsystem Aa,Ab  was detected
from  the wavy relative  motion of  A,B \citep{Hei1996b}  and directly
resolved  for the  first time  by  speckle interferometry  in 1982  as
CHR~6. In 2002,  star C was also resolved into a 0\farcs4 pair Ca,Cb
(CTU  2).  The  period  of  Ca,Cb estimated  from  its  separation  is
$\sim$60\,yr and its orbit is not known yet.

Gaia measured  concordant parallaxes of A,  B, and C; the latter,
most accurate (22.22$\pm$0.08\,mas), is adopted here.  The orbit of A,B
with  $P=620$\,yr  \citep{Hei1996b}, updated  here  to $P  =2400$\,yr,
remains  essentially unconstrained.  \citet{Drummond2003}  reached the
same  conclusion and  modeled the  observed segment  of this  orbit by
a linear  motion. They combined  the wobble  with direct  resolutions of
Aa,Ab,  computed its  first orbit  with  $P =  47$\,yr, and  estimated
masses  of  Aa  and  Ab;  their results are refined  here (Figure~\ref{fig:02291}).   

The orbit of Aa,Ab is now very well constrained and, together with the
Gaia parallax, defines  the mass sum of Aa,Ab,  2.91 \msun. The wobble
factor $f = 0.322 \pm 0.013$  implies the mass ratio of $q_{\rm Aa,Ab}
=  0.47$, hence  the masses  of Aa  and Ab  are 1.98  and  0.93 \msun,
respectively.  The  absolute magnitudes of main-sequence  stars Aa and
Ab (visual magnitudes 4.65 and 8.63 mag) agree with these masses.  The
mass of B estimated from its  magnitude is 1.28 \msun, hence the outer
mass sum should be 4.2 \msun.

Comparable  separations  between A,  B,  and  C  raise concerns  about
dynamical  stability of this  system and  thus constrain  somewhat the
orbit of A,B. On the one hand,  its semimajor axis must not be too big
compared to the 7\arcsec   ~projected separation between A and C,
on the  other hand  the periastron separation  must be at  least three
times  larger than  the semimajor  axis of  Aa,Ab,  i.e.  $>$1\farcs3,
which rules out large eccentricities.  I fixed the eccentricity of A,B
at 0.4  and also  fixed the period  (2400\,yr) and the  semimajor axis
(6\farcs5)  to values  that yield  the expected  mass sum,  4.2 \msun.
Without these restrictions, the errors of  $P, e, a$ from the free fit
are huge.  The tentative orbit of A,B proposed by \citet{Hei1996b} has
a smaller semimajor  axis of 2\farcs9 and a  shorter period and yields
the mass sum of 5.9 \msun.
 
The 6\farcs5  semimajor axis  of A,B is  in tension with  the 7\arcsec
 ~projected separation between A  and C. A circular orbit of AB,C
with  $P^*  \sim 2$\,kyr,  estimated  from  the projected  separation,
corresponds to the  relative orbital speed of $\mu^*  = 18$ \masyr.  I
computed the observed speed of  the relative motion between AB (center
of mass)  and C  using historic measurements  of A,C,  subtracting the
orbital motion of  A, and accounting for the  precession in angle. The
resulting relative motion speed is  only 4.1 \masyr. It is directed at
an angle  of 31\degr ~to the  vector AB,C (C approaches  AB). The slow
relative motion is a sign that  the actual separation between AB and C
 in  space is substantially larger than  the projected separation.
    In   such   case,   the   large  semimajor   axis   of   A,B   is
plausible. However,  even if the ratio  of  the  semimajor axes of
  AB,C  and A,B  is compatible  with the  dynamical stability,  it is
still moderate, suggesting dynamical evolution of this system.

Although the orbit of A,B is poorly constrained, the relative
inclination between A,B and Aa,Ab is almost insensitive to the choice
of $P,e,a$ and has values of either 46\degr ~or 108\degr. Kozai-Lidov
cycles \citep{Naoz2016} in the inner triple are expected, and the large inner
eccentricity of 0.64 may be a manifestation of these cycles. 

\subsection{04375$+$1509 (vB 102)}
  
\begin{figure}
\epsscale{1.0}
\plotone{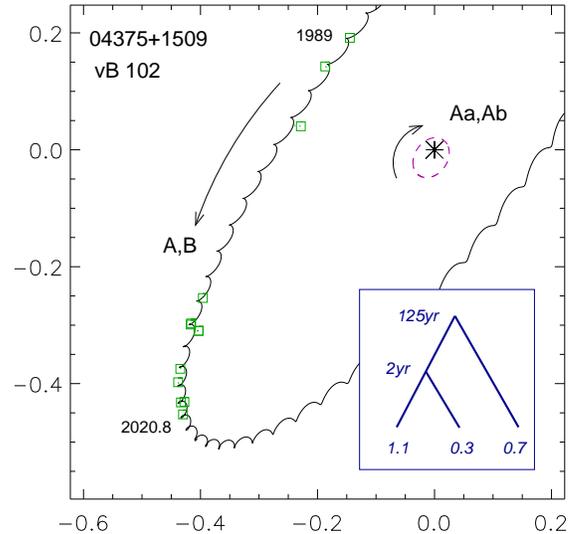}
\caption{Outer orbit of vB~102 (125 yr) with a 2-yr wobble. The inner
  pair Aa,Ab remains unresolved. 
\label{fig:04375}  }
\end{figure}

This triple system is a G0V star in the Hyades cluster, often referred
to as vB~102. It is  a single-lined spectroscopic binary with a period
of  734.2  day  \citep{Griffin2012}  with  a  faint  visual  companion
discovered in 1989 at  0\farcs25 separation and designated as CHR~153.
Thirty years of interferometric  coverage resulted in the first, still
tentative  visual orbit  with  $P=128$ yr  \citep{Tok2019c}. The  Gaia
parallax of  20.61$\pm$0.46 mas  is used here,  although it  is likely
biased by the multiplicity; the PM anomaly is large \citep{Brandt2018}.

The estimated inner semimajor axis is 37\,mas, but the inner subsystem
remains unresolved owing to the large contrast; however, the wobble in the
observed motion of the outer pair is detectable.  Two orbits fitted to
the  speckle  measurements  and  RVs (Figure~\ref{fig:04375})  have  a
wobble amplitude of 8.5\,mas, or  $f=0.22 \pm 0.04$ and $q_{\rm Aa,Ab}
= 0.29$ (the  inner semimajor axis is computed  from the third Kepler's
law).  The mass  of the G0V star Aa is estimated at  1.13 \msun, hence the
mass of  Ab is  about 0.32 \msun.   Most measurements come  from SOAR.
\citet{Griffin2012} determined the negative  linear RV trend caused by
the  outer  system and  suggested  that  $\omega_{\rm  A,B}$ is  about
270\degr, as is  indeed the case. Note that the  RV coverage starts in
1959; it exceeds the speckle coverage and helps to constrain the outer
orbit. In  the final  fit I  fixed the outer  eccentricity at  0.2 and
obtained  the expected outer  mass sum  of 2.16  \msun.  The  outer RV
amplitude   is  2.23$\pm$0.23   \kms  and   the  system   velocity  is
$41.57\pm0.41$ \kms.

The  RV  amplitude  in  the  inner  orbit  is  small,  3.85  \kms,  and
corresponds    to   the    minimum    mass   of    0.17   \msun    for
Ab.  \citet{Bender2008} apparently detected  weak lines  of Ab  in the
infrared spectra and deduced the Ab  mass of 0.17 \msun ~from the three
RVs, implying a  large inclination of the inner  orbit. However, their
RVs deviate  from the  orbit more than  allowed by the  claimed errors
(rms 6.6 \kms), and Griffin considered them as spurious.  The wobble
amplitude measured here also disagrees with the infrared RVs.

The most unusual result is  the apparent counter-rotation in the inner
and  outer orbits  (Figure~\ref{fig:04375}).  The free   fit
gives  the  inner  inclination   of  $i_{\rm  Aa,Ab}  =  153\degr  \pm
60\degr$. I fixed the inclination to 144\degr ~for consistency between
wobble  and RV  amplitudes; both  correspond to  the Ab  mass  of 0.33
\msun.  Attempts to  fit a  co-rotating inner  orbit fail.  Mutual
inclination between the orbits is 72\degr ~with fixed inner inclination
or $82\degr \pm 47 \degr$ with the free fit (both nodes are known).

Admittedly, the wobble signal is weak and the astrometric inner orbit
is still preliminary. Future Gaia data releases will define this orbit
much better if the triple nature of this system can be properly
accounted for in the data reduction of this mission.

\subsection{04397$+$0952 (GJ 173.1)}

\begin{figure}
\epsscale{1.0}
\plotone{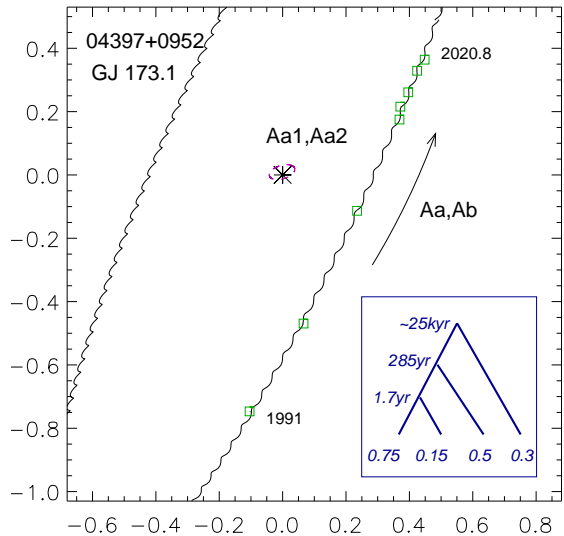}
\caption{Outer orbit of GJ 173.1 Aa,Ab ($P=285$ yr) with a 1.7 yr wobble. 
\label{fig:04397}  }
\end{figure}

This nearby  (29\,pc) quadruple system  is composed of  low-mass stars
(Figure~\ref{fig:04397}).   The  primary  (GJ~173.1,  G 83-28)  has  a
spectral type K0V.   The 14th magnitude star B  (G 83-29) at 34\arcsec
~has  a  common PM.   The  Gaia  parallax  of B,  34.04$\pm$0.12\,mas,
defines the distance to the  system better than the biased parallax of
A, 36.80$\pm$0.39\,mas.  Another visual companion Ab,  3.4 mag fainter
and  at 0\farcs75  separation,  was detected  by Hipparcos  (HDS~601).
Most observations of this pair  Aa,Ab come from the speckle program at
SOAR. Star Aa is also a single-lined spectroscopic binary Aa1,Aa2 with
$P=610.43$\,day  \citep{Halbwachs2018,Sperauskas2019}.   The estimated
semimajor axis  of the innermost  orbit, 46\,mas, favors  detection of
the wobble signal.

The orbit of Aa,Ab, including  wobble, was fitted to the eight available
position measurements  of Aa,Ab and  40 published RVs of  Aa1 covering
the  period from  1986  to 2014  (Figure~\ref{fig:04397}).  The  outer
orbit is not fully constrained, and  its period is fixed to 285\,yr to
yield  the  expected  outer  mass  sum of  1.4  \msun;  the  published
preliminary  orbit  had $P=204$\,yr  \citep{Tok2017c}. Even with the
fixed period, the errors  of some  elements are large.  A negative  RV
trend caused  by the  outer orbit  is obvious ($K_3  = 1.9  \pm 0.6$
\kms),  fixing its  correct node.  The  system velocity  is $\gamma  =
-26.2 \pm 1.0$ \kms.

The wobble factor $f =  0.17 \pm 0.03$ corresponds to $q_{\rm Aa1,Aa2}
= 0.20$, so  the estimated mass of Aa1, 0.75 \msun,  leads to the mass
of 0.15  \msun for Aa2.  The RV  amplitude $K_1 = 4.47  \pm 0.10$ \kms
corresponds  to  the secondary  mass  of  0.17  \msun, hence  the  two
estimates  match within  errors.    The mass of  the pair Aa is
0.90 \msun, and the outer RV  amplitude implies the mass of 0.53 \msun
for  Ab   The absolute magnitude of Ab corresponds to a star of 0.48
\msun, in agreement with its spectroscopically estimated mass.

This system resembles the previous one. The orbital nodes are known,
and the mutual inclination deduced from the wobble is large,
51\degr$\pm$10\degr. This result is a consequence of the different node
angles $\Omega$, while the inclinations of the inner and outer orbits
are  similar (both are large). An attempt to force approximate orbit coplanarity
fails. The weighted rms residuals of positional measurements are
$\sim$1\,mas if the wobble is fitted and 5\,mas without wobble. 

\subsection{04400$+$5328 (ADS 3358)}

\begin{figure}
\epsscale{1.0}
\plotone{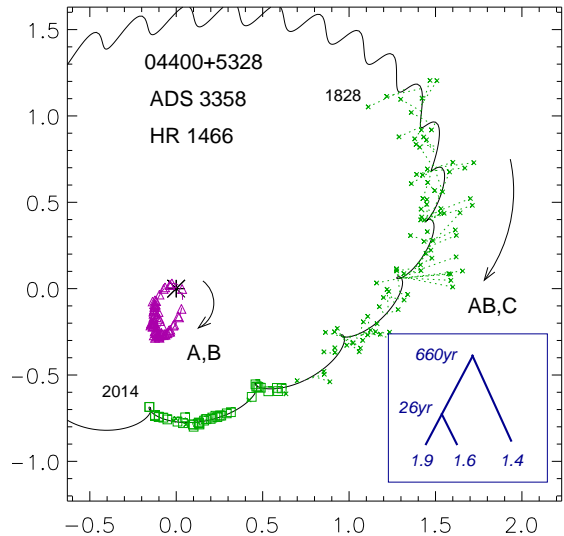}
\caption{Outer (STF 566, 660 yr) and inner (BU 1295, 26 yr) orbits of ADS 3358. 
\label{fig:04400}  }
\end{figure}

The classical resolved visual  triple system 2~Cam (HR~1466, ADS 3358,
spectral  type   A8V)  has  been   the  subject  of   several  studies
\citep[e.g.][]{Hei1996b}.  Here  both orbits are  updated using recent
speckle measurements (Figure~\ref{fig:04400}).  The eccentric orbit of
the inner  26 yr pair  A,B (BU~1295) is  now fully covered  by the speckle
data  and of excellent  quality (the weighted  rms residuals  are 5\,mas).
Several  measurements  of the visual  magnitude  difference  in this  pair
average at $\Delta V_{\rm A,B} =  1.29$ mag. I adopt the masses of 1.94
and 1.45 \msun deduced from  the absolute magnitudes of these stars and
the  dynamical parallax of  13.0\,mas.  The Gaia  parallax of
15.3$\pm$0.4 mas is obviously inaccurate and biased (in 2015 the inner
pair,  unresolved by  Gaia, moved  rapidly through  the  periastron at
0\farcs07 separation) and implies an unrealistically small inner mass sum
of 2.1 \msun.

The outer orbit of AB,C (STF  566) is not so well defined, despite its
numerous  measurements  available since  1828  that  cover a  144\degr
~arc. I  fixed the  outer period to  660\,yr and fitted  the remaining
elements with  wobble.  The  wobble amplitude is  very accurate  ($f =
0.446 \pm 0.009$) and the measured mass ratio $q_{\rm A,B} = 0.805$ is
close to  the ratio  of the adopted  masses, $q_{\rm  A,B}=0.77$.  The
outer  mass sum  computed with  the dynamical  parallax is  4.8 \msun,
approximately  matching the  mass  of C  estimated  from its  absolute
magnitude.

Two possible values of mutual inclination deduced from the orbits are
close to each other: 57\fdg3$\pm$2\fdg2 or 61\fdg7. Therefore, the
orbits are highly inclined. Note the large inner eccentricity of 0.85
suggestive of Kozai-Lidov cycles in this system. 

\subsection{15440$+$0231 (23 Ser, GJ 596.1)}

\begin{figure}
\epsscale{1.0}
\plotone{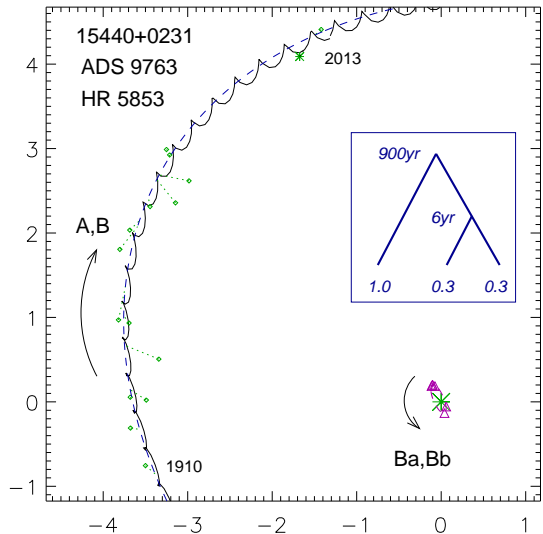}
\caption{Orbits of GJ 596.1 (ADS 9763), periods 900 yr and 6.6 yr.
\label{fig:15440}  }
\end{figure}

The bright star  23~Ser (HR 5853, GJ 596.1,  ADS~9763), located within
15 pc from the Sun, has  been known to host a faint physical satellite
B since 1910, when this pair was first measured by R.~Aitken (A~2230).
The latest orbit of A,B with $P=529$\,yr and $a=5\farcs04$ computed by
\citet{Gat2013b}  is only  preliminary,  despite the  observed arc  of
85\degr.   Other companions  mentioned in  the WDS  are  optical.  The
bright and  nearby G3V  star A has  been studied from  various angles,
including non-detection of planets.

The  M3V  star   B  was  resolved  into  a   0\farcs2  pair  Ba,Bb  by
\citet{RDR2015} in  2009 and  re-discovered at approximately  the same
position in  2015 by \citet{Tok2016d}.  They suggested  that Ba,Bb has
made one full revolution between  these observations and its period is
about 6 yr.  Three additional  observations made at SOAR (the last one
in  2020.1)  now fully  constrain  this  orbit  with $P=6.6$\,yr  and
$a=0\farcs19$. The Gaia parallax of star A, 67.71$\pm$0.07\,mas, leads to
the inner mass sum of  0.50 \msun. The magnitude difference between Ba
and Bb measured  by \citet{Tok2016d}, 0.3 mag, tells  us that stars Ba
and  Bb are  similar  (a twin),  so  their masses  are  0.26 and  0.24
\msun. The photometry by  \citet{RDR2015} defines the $K$ magnitude of
Ba, 8.08 mag, which corresponds to the mass of 0.27 \msun according to
the standard relations.

The estimated  mass of A,  0.98 \msun, implies  the outer mass  sum of
1.48  \msun. I fixed  the outer elements  $P=900$\,yr and  $a=7\farcs2$ to
match    this    mass    sum    and    fitted the   remaining    elements
(Figure~\ref{fig:15440}). The  orbital  motion  changes  the   PM  of
A. According to  \citet{Brandt2018}, the Gaia PM anomaly  of star A is
$(+1.3, +3.8)$ \masyr.   The difference between the orbital speed  of B in
2015.5 and its mean speed between 1991.25 and 2015.5 is $(-3.1, -7.2)$
\masyr,  in  qualitative  agreement  with  the  PM  anomaly.   The  RV
amplitude in the outer orbit is $\sim$1 \kms, explaining the linear RV
trend of $-3.04$ m~s$^{-1}$ per year detected by \citet{Butler2017}.

Only one resolved  measurement of A,Ba is available,  hence the wobble
amplitude  remains unknown; I  fixed it  to $f  = -0.5$.  Opposite
rotation  directions in  the inner  and outer  pairs are  evident. The
relative inclinations are 75\degr$\pm$3\degr ~or 141\degr. This triple
system is definitely not co-planar.

\subsection{16057$-$0617 (ADS 9918) }

\begin{figure}
\epsscale{1.0}
\plotone{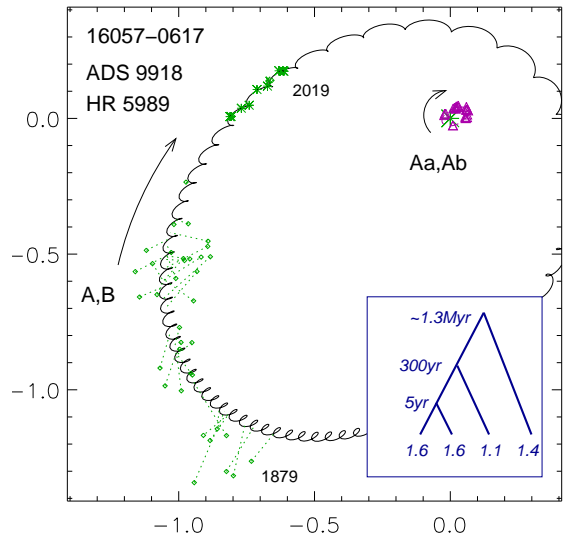}
\caption{Fragment of the the outer orbit (300 yr) of ADS 9918. The
  inner period of Aa,Ab is 5 yr.
\label{fig:16057}  }
\end{figure} 

This is a classical visual triple  system (ADS 9918, HR 5989) with two
known orbits and the  fourth distant (260\arcsec) and bright component
E  (HIP  78822, HD  144309)  with common  PM  and  parallax. The  Gaia
parallax of E,  12.03$\pm$0.05\,mas, defines the  distance better  than the
less accurate and potentially biased parallax of A, 12.2$\pm$0.3\,mas.

The  outer  pair  A,B  (BU   948),  discovered  in  1879  at  1\farcs5
separation,  is  slowly closing  down;  the  observed  arc is  75\degr
~(Figure~\ref{fig:16057}).  The  inner  subsystem Aa,Ab  discovered  by
W.~Finsen in 1964 (FIN 384) has an orbital period of exactly 5 yr. Components Aa
and  Ab are  equal and  their separation  never exceeds  70\,mas.  The
orbit    of    Aa,Ab    has     a    good    coverage    by    speckle
interferometry. Measurements of Aa,Ab at close  separations in 2008  at SOAR
and in 2017  at Gemini \citep{Horch2019},  at phases not  covered before,
further constrain the  orbit. The inner mass sum  of 3.26 \msun
matches the
absolute magnitude of its equal components of spectral type F2IV.

The  outer  orbit  with  $P=459$\,yr  and  $e=0.65$  was  computed  by
\citet{Zir2014a}. The data do not  constrain the period, so I fixed it
to 300  yr and, furthermore,  fixed the outer inclination  to 155\degr
~to obtain the expected outer mass  sum, 4.3 \msun (the free fit gives
$i_{\rm  AB}  =  158\degr  \pm  11\degr$).  The  wobble  factor  $f  =
0.49\pm0.05$ confirms  that the masses of  Aa and Ab are  equal.  However,
these  stars  were  sometimes  swapped  and  some  outer  measurements
referred  to   Ab,B  instead  of  Aa,B;  these   cases  are  rectified
here. Moreover, the  measurement of Aa,B at Gemini  in 2017.4 revealed
inaccurate calibration of the scale  and was given a low weight, while
the   tight  inner  pair   Aa,Ab  was   measured  well   at  27.5\,mas
separation. The  mutual inclination between the orbits  is 32\degr ~or
78\degr; the smaller inclination appears more likely.
 
\subsection{17066$+$0039 (ADS 10341)}

\begin{figure}
\epsscale{1.0}
\plotone{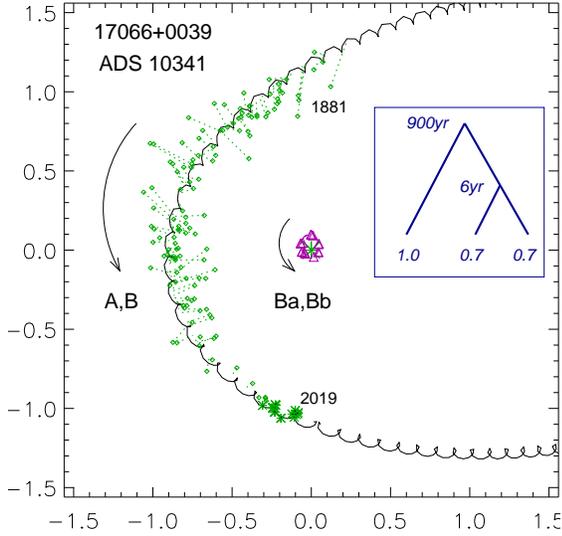}
\caption{The outer orbit (900 yr) with wobble and the inner 6.3 yr orbit of ADS 10341.
\label{fig:17066}  }
\end{figure}

The outer pair of ADS~10341 (BU  823) is known since 1881.  The latest
visual orbit  of A,B with $P=532$\,yr computed  by \citet{Hrt2000c} is
still preliminary, although the observed arc slightly exceeds 180\degr
(Figure~\ref{fig:17066}).  The  main G0V star A has  an estimated mass
of 0.98 \msun and its  Gaia parallax is 17.90$\pm$0.07 mas.  

The secondary component B  was resolved at SOAR in 2009  into a close pair
Ba,Bb  (TOK~52).   The  orbit  of  Ba,Bb  with   $P=6.3$\,yr  is  well
constrained \citep{Tok2016}.   It is  slightly updated here  using the
latest measurements  \citep[e.g.][]{Horch2019}. The mass  sum of Ba,Bb,
 1.50 \msun,  agrees  with  masses estimated  from the  absolute
magnitudes of Ba  and Bb.  The magnitude difference  between Ba and Bb
is small,  $\sim$0.1 mag, and the  wobble factor $f =  -0.52 \pm 0.03$
confirms the equality  of masses.  No wobble is  therefore expected in
the positions  of A,B, and only  the resolved measurements  of A,Ba at
SOAR show characteristic loops with a 6 yr period.

The period of A,B is between 350 yr and 1000 yr.  The curvature of the
observed  arc constrains  the mass  sum to  values between  2  and 2.5
\msun. The  period in the free fit  has a large error,
hence it can be selected  within the error range.  Orbits with shorter
periods have smaller eccentricities  and inclinations and the mass sum
around  2 \msun.  I  fix $P_{\rm  A,B} =  900$ yr  and $i_{\rm  A,B} =
40\degr$ to  obtain the mass  sum of 2.5  \msun, in agreement  with the
estimated mass of A.

The two  possible values of  mutual inclination, 60\degr  ~or 46\degr,
are  similar.  If a  shorter  outer  period  is enforced,  the  mutual
inclination is  reduced to $\sim$33\degr.  The inner  and outer orbits
are thus aligned, but only approximately so.

\subsection{17157$-$0949 (ADS 10423)} 

\begin{figure}
\epsscale{1.0}
\plotone{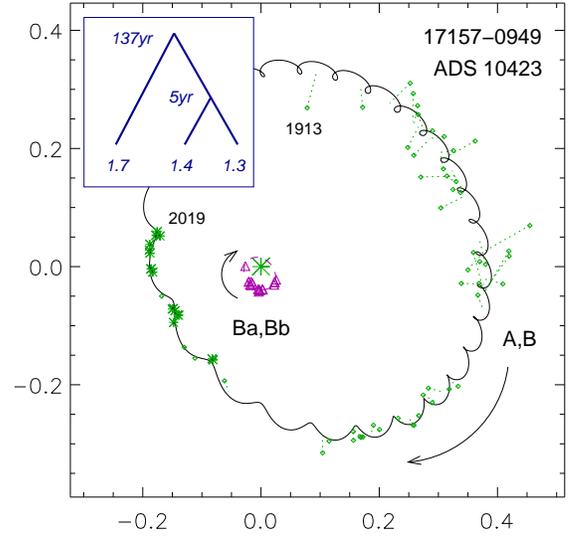}
\caption{Orbits of ADS 10423 (137 yr and 5.26 yr).
\label{fig:17157}  }
\end{figure}

\begin{figure}
\epsscale{1.0}
\plotone{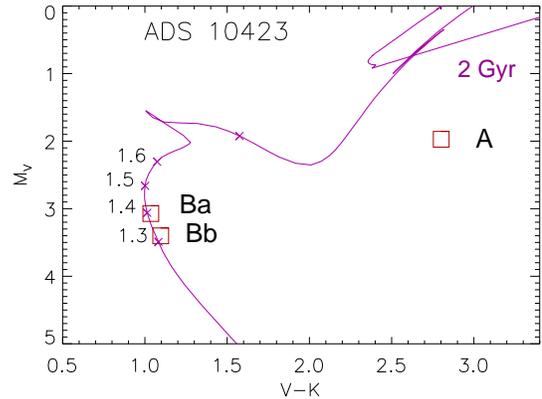}
\caption{Components of ADS 10423 in  the $V,V-K$ CMD (squares) and the
  2 Gyr  isochrone for  solar metallicity \citep{PARSEC}.  Numbers and
  crosses mark masses from 1.3 to 1.7 \msun ~on the isochrone.
\label{fig:A2592}  }
\end{figure}

The  7th mag star HD~156034 was resolved by  R.~Aitken in 1913 as a 0\farcs3 pair
with components of  similar brightness (ADS 10423, A  2592). Its orbit
has been  calculated and  re-calculated many times.   Now most  of the
outer orbit  with $P=137$\,yr  is covered and  its elements  are quite
accurate (Figure~\ref{fig:17157}).

The inner subsystem  Ba,Bb (TOK 53) was discovered at  SOAR in 2009. Its
preliminary  5 yr  orbit \citep{Tok2015c}  is  well defined  now, after  two
complete  revolutions have been  monitored at  SOAR. The  weighted rms
residuals of  Ba,Bb positions are  1.5 mas. The latest  measurement of
Ba,Bb  made in  2020.2 at  27\,mas separation  (below  the diffraction
limit) has a large residual in angle  and is assigned a low weight. The
inner semimajor axis  is 33 mas and the Ba,Bb pair  is not resolved at
SOAR near periastron. The wobble  amplitude is 14.5\,mas ($f = -0.44
\pm 0.04$).

This triple  system is lucky  in having both orbits  well constrained,
but its  analysis is nevertheless  tricky because the distance  is not
accurately measured  and the components are evolved.  The original and
revised     Hipparcos    parallaxes    are     5.7$\pm$1.3\,mas    and
4.9$\pm$0.9\,mas,  respectively.   Gaia  normally  does   not  measure
parallaxes and PMs of close binaries (A,B was at 0\farcs18 in 2015.5),
but it did so for ADS  10423; the Gaia parallax of 2.8$\pm$0.9\,mas is
obviously  erroneous.   All  three  quoted   trigonometric  parallaxes
correspond  to unrealistically  large masses.   I adopt  the dynamical
parallax of 7.9\,mas based on the orbits and estimated masses.

The spectral  type F5V  quoted in Simbad  disagrees strongly  with the
combined color $V-K  = 2.40$ mag that corresponds  to the spectral type
K3,  matching the low  effective temperature  of 4933\,K  estimated by
Gaia.   This contradiction  suggests that  star A  is redder  and more
evolved compared to  the less massive stars Ba and Bb.  The light of B
dominates  at short  wavelengths and  its estimated  spectral  type is
indeed F5V.  I propose  the system model  where star  A has a  mass of
1.705 \msun  and stars Ba and Bb  have masses of 1.38  and 1.30 \msun,
respectively. The age of the system is about 2 Gyr.

Figure~\ref{fig:A2592} compares absolute  magnitudes and colors of the
components  with the  2  Gyr PARSEC  isochrone  for solar  metallicity
\citep{PARSEC}.  However, the $V-K$ colors of stars are guessed rather
than measured  because no differential  photometry in the $K$  band is
available. I  use the magnitude difference  between Ba and  Bb of 0.32
and 0.30  mag measured by \citet{Horch2019} at  wavelengths of 692\,nm
and  880\,nm  (roughly  in  the   $R$  and  $I$  bands)  and  confirmed
independently by the differential speckle photometry at SOAR, e.g. $\Delta
y_{\rm  Ba,Bb}  =  0.33$  mag.  According  to  \citet{Horch2019},  the
magnitude difference between A and Ba  is 1.30 and 1.66 mag in the $R$
and $I$ bands,  and SOAR complements this with  $\Delta y_{\rm A,Ba} =
1.11$ mag.  Photometry confirms that Ba is bluer than A. The two-color
photometry of A and B provided by Tycho is discrepant and I ignore it,
while  the Hipparcos  measurement  $\Delta Hp_{\rm  A,B}  = 0.27$  mag
appears reliable.   The $V$ magnitudes of  A, Ba, and  Bb deduced from
the combined  and differential  photometry are  7.51, 8.61,  and  8.94 mag,
respectively. I assign the $K$ magnitudes  of Ba and Bb to match their
positions on the isochrone (and  the spectral type F5), and obtain the
$K$ magnitude of  A from the combined photometry;  the result is 4.71,
7.57, and 7.85 mag.

The proposed system model defines  the masses of the components quoted
above  and matches  the differential  photometry  in the  $V$ and  $I$
bands.  The outer mass sum of 4.4 \msun deduced from the model and the
outer  orbit correspond to  the dynamical  parallax of  7.9\,mas. With
this parallax, the less accurate inner orbit gives the mass sum of 2.6
\msun  for Ba,Bb, in  good agreement  with the  model. The  model thus
matches  both the  orbits and  the  photometry. The  inner mass  ratio
$q_{\rm Ba,Bb}  = 0.93$ corresponds to $f=0.48$,  which agrees, within
errors,  with  the measured  wobble  factor  $f=0.44$.  If  additional
multi-color photometry and high-resolution spectroscopy are secured in
the  future  and  properly  interpreted,  this system  might  offer  a
valuable test  of the isochrones because  Aa is just  leaving the main
sequence.

Mutual  inclination between the orbits  is either  29\degr $\pm$14\degr  or
77\degr; the  first, smallest value  seems more likely.  It  is worth
mentioning  that the  RV of  this  star (apparently  dominated by  the
late-type spectrum of A) has  been monitored by \citet{TS2002} and found
to be constant.
 
\subsection{19453$-$6823 (HIP 97196)}

\begin{figure}
\plotone{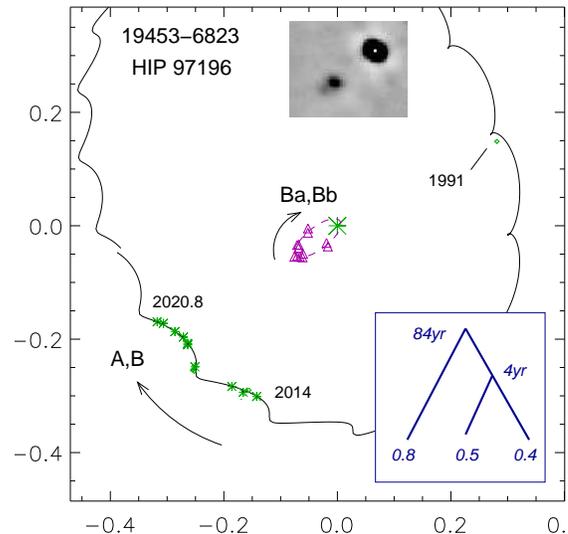}
\caption{Orbits of HIP~97196 (84 and 4.4 yr). The insert shows the speckle
ACF recorded in 2018.8.
\label{fig:19453}
}
\end{figure}

An inconspicuous 10th  magnitude K-type dwarf star HIP  97196 has been
resolved by  Hipparcos into  a 0\farcs3 pair  (HDS 2806)  with unequal
components, $\Delta Hp = 2.63$  mag. The binary has been observed for
the second  time at  SOAR in 2014.3,  and its secondary  component was
resolved into a  0\farcs05 pair Ba,Bb (TOK~425) with  $\Delta I = 1.2$
mag. As expected, the inner pair  is a fast mover, and its first orbit
with  $P= 4.1$  yr was  published by  \citet{Tok2019c}. The  insert in
Figure~\ref{fig:19453}   shows  a  typical   speckle  auto-correlation
function (ACF) to illustrate the difficulty of measuring this tight and
faint  triple system  with substantial  contrast. All  observations at
SOAR were made in the $I$ filter, the last one in 2020.8.

About half of the outer orbit is now covered, and the first set of its
elements is  computed here.  However, the coverage  is uneven,  with a
large gap between  1991 and 2014. The orbit  is not fully constrained:
shorter periods with  smaller eccentricity also fit the  data. I fixed
the outer inclination to 140\degr ~to match the expected mass sum. The
eccentric  ($e_{\rm  Ba,Bb}  =0.85$)  inner  orbit is  also  not  very
accurate because its part near  the periastron is below the resolution
limit.   The  quadrants   of   Ba,Bb  are   defined,  so the  alternative
low-eccentricity orbit with double period is not allowed.  The closest
resolved measurement of Ba,Bb at 36\,mas separation was made in 2017.4
at  Gemini  \citep{Horch2019},   at  880\,nm  wavelength. Their simultaneous
measurement at  692\,nm has discordant  photometry and is given  a low
weight, so this object is a difficult one even for an 8 m telescope. 

The  Gaia parallax  of 21.02$\pm$0.39  mas  is not  very accurate  and
possibly  biased,  although  it  agrees  with  the  revised  Hipparcos
parallax of 21.3$\pm$1.9\,mas \citep{HIP2}.  I adopt the Gaia parallax
and  estimate masses  from  the standard  relations  and absolute  $I$
magnitudes. The combined $I=8.83$ mag is deduced by interpolation from
other bands (no direct measurement of $I$ is found in the literature),
and I adopt $\Delta I_{\rm A,B} = 2.0$ mag and $\Delta I_{\rm Ba,Bb} =
1.2$ mag.  The estimated masses of A,  Ba, and Bb are  0.78, 0.54, and
0.40 \msun, respectively.  Main-sequence stars with these masses match
the combined  magnitudes in  the $V$ and  $K$ bands and  the magnitude
differences,  the orbits, and  the Gaia  parallax.  The  model implies
an inner  mass ratio of $q_{\rm Ba,Bb}  =  0.74$ and  $f=-0.43$, while  the
measured wobble factor is $f = -0.34 \pm 0.03$.

The  outer orbit matches  the PM  anomaly of  $(17.8, 7.6)$  \masyr in
1991.25  determined by  \citet{Brandt2018}  because the  corresponding
orbital  motion  of  B  relative  to  A,  $(-31.3,-20.5)$  \masyr,  has
approximately  opposite  direction  and  is  two  times  faster.   The
photo-center  wobble factor  $f^*$ in  the  outer orbit in the $V$  band,
computed  using the  adopted masses,  is  0.47.  The  Gaia PM  anomaly,
$(-1.2, -19.2)$ \masyr, does not agree so well with the orbital motion
of AB, $(9.3, 37.73)$ \masyr, possibly because of the shorter time span
of  Gaia data  and  the larger  effect  of the  inner  subsystem on  the
photo-center motion in the $G$ band. Astrometric signal caused by both
orbits should be recovered correctly in the future Gaia data releases
that will also yield the unbiased parallax.

Both inner and outer subsystems  of HIP~97196 rotate clockwise, but the
relative inclination  is substantial, $\Phi  = 42\degr \pm  7\degr$ or
$\Phi =  80\degr$; the  period ratio is  about 20. The  current linear
apparent  configuration of  these stars on the sky  suggests coplanar  and highly
inclined orbits, but this impression is misleading.  It is likely that
this triple system goes through Kozai-Lidov cycles and currently is in
the low-inclination  and high-$e$ phase of the  cycle. Evidently, this
hierarchical system differs  from the quasi-coplanar ``dancing twins''
with  comparably  small  masses  and short  periods  \citep{twins}  in
several  important  respects:  large  mutual inclination,  large  inner
eccentricity, and unequal masses of the components.

The  spatial velocity  of  this system  $U,V,W  = (45,  4, -46)$  \kms
strongly suggests that  it is old and possibly  metal-poor. Its further
detailed  study  will be  interesting  for  various  reasons, e.g.  to
measure  accurate  masses  and  luminosities and  to  probe the long-term
dynamical evolution. Continued speckle monitoring, relative photometry
in the infra-red, and high-resolution spectroscopy are needed.

\subsection{22300$+$0426 (ADS 15988)}

\begin{figure}
\plotone{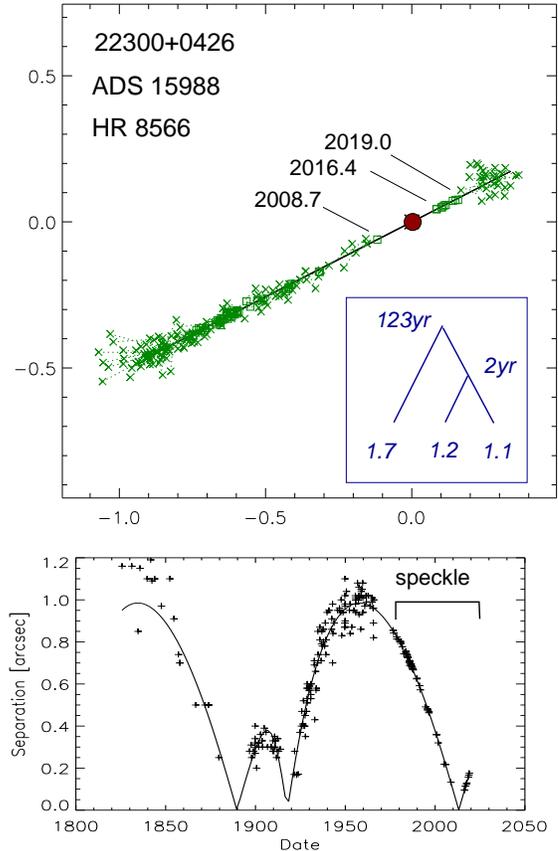}
\caption{Visual  orbit of the outer subsystem in ADS 15988 (STF 2192).
  Top: orbit on the sky, bottom: separation vs. time. 
\label{fig:A}
}
\end{figure}

The bright star 37 Peg (HR 8566, ADS 15988) was resolved as a 1\arcsec
~visual pair by W.~Struve in 1825 and is known as STF~2192.  This pair
slowly  closed  down  during   the  19th  century,  went  through  the
periastron around 1911,  opened up again and started  to close down in
the  2nd half of  the 20th  century.  From  1976 on,  accurate speckle
interferometry  replaced  visual  micrometer  measurements,  providing
excellent coverage of the  decreasing angular separation.  This bright
pair was frequently observed  by speckle interferometry, possibly as a
convenient  calibrator.  Its  latest  orbit by  \citet{Sod1999} has  a
grade 2,  i.e.  is  reliable and accurate  by current  standards.  The
inclination is  90\degr, so  the orbit is  based only on  the measured
separations  (Figure~\ref{fig:A}).   The  WDS  database  contains  335
measurements of its relative position.

The   parallax   of   37    Peg   was   measured   by   Hipparcos   at
18.93$\pm$1.23\,mas,  revised to 19.1\,mas  by \citet{Sod1999}  and to
19.58$\pm$0.58  by  \citet{HIP2}.   The Gaia  parallax of 18.84$\pm$0.39
(53\,pc  distance) is  only slightly  more accurate,  possibly because
this  star is bright  and its  astrometry is  disturbed by  the orbital
acceleration.  Fortunately, motion in the outer orbit between 1991 and
2015 was almost linear and the PM anomaly is small \citep{Brandt2018}.
All parallax measurements are mutually consistent, within their errors.
The  orbital elements  and the masses estimated  here  favor a  slightly
smaller dynamical parallax of 18.4\,mas.

This $V=5.5$ mag star  attracted attention of  observers for
various    reasons.    It    was   monitored    spectroscopically   by
\citet{Abt1976} in a survey of  bright solar-type stars.  They found a
spectroscopic   subsystem  with   a  period   of  372.4   day   and  a
semi-amplitude of  8.2 \kms. However, this orbit,  together with other
orbits  from  that  paper,   has  not  been  confirmed  by  subsequent
observations.  The star is no  longer listed as a spectroscopic binary
in the current catalog. And yet it is triple!

\begin{figure}
\plotone{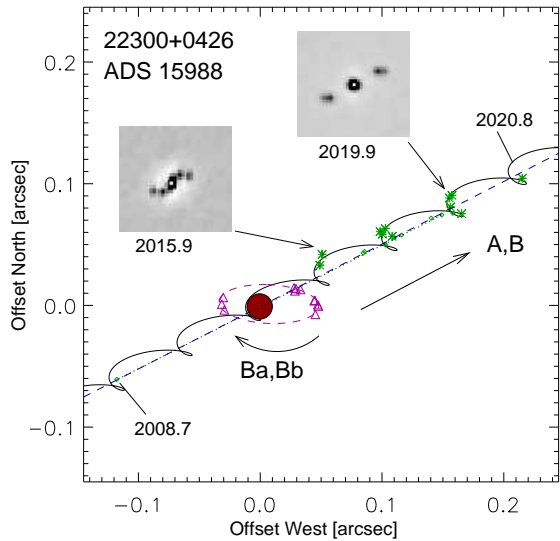}
\caption{Fragment of the outer orbit of ADS 15988 covered at SOAR. The
  wavy line is the position of Ba relative to A (the red circle at the
  coordinate origin) affected by the wobble, asterisks show the
  measured positions of Ba. Small crosses and the blue dash line show the
  unresolved measurements of B without wobble.  The inner orbit is
  over-plotted on the same scale (the magenta  ellipse and triangles). 
\label{fig:wobble}
}
\end{figure}

ADS 15988 was observed by speckle interferometry at SOAR for the first
time in 2008.7 when it still approached the conjunction. Shortly after
the conjunction, in  2015, the pair was re-visited  at SOAR twice.  On
the second  visit in  2015.9,  star B was  clearly resolved  into a
close  46 mas  subsystem (Figure~\ref{fig:wobble})  with  nearly equal
components  Ba and  Bb \citep{Tok2016b}.  We  can only  guess why  the
subsystem Ba,Bb  has not been  noted much earlier in  numerous speckle
observations with  4 m telescopes.  Maybe the  fact that it  is such a
well-known binary  prevented observers from noting  that the secondary
component is itself a close pair.

The small separation of Ba,Bb implied a short orbital period. For this
reason,  the object was  frequently observed  at SOAR.   The subsystem
played  ``hide and  seek'': it  disappeared in  2016 but  was resolved
again in 2017--2020. In 2017,  the subsystem was also measured with the
8  m Gemini  telescope \citep{Horch2019}.   Finally,  the measurements
accumulated to  date allow  calculation of the  orbit of Ba,Bb  with a
period  of  2.1  yr  (Figure~\ref{fig:wobble}).  The  orbit  of  Ba,Bb
successfully models  the measurements (the weighted  rms residuals are
0.9\,mas)  and  predicts small  separations  for  the  dates when  the
subsystem was  not resolved, e.g.  in 2008.7. The  observations cover two
orbital periods  of Ba,Bb.   The Gaia parallax  of 18.84$\pm$0.39\,mas
corresponds  to the  inner  mass sum  of  2.34$\pm$0.25 \msun. 

\begin{figure}
\plotone{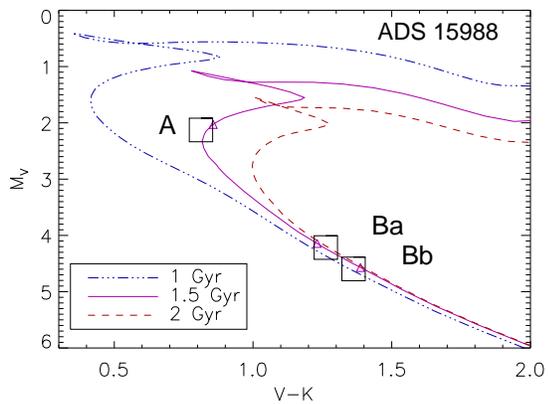}
\caption{Location  of  the components  of  ADS 15988  (squares) on  the
  color-magnitude   diagram.    The   lines  are   PARSEC   isochrones
  \citep{PARSEC}  for  solar  metallicity.  Small triangles  mark  the
  masses of 1.70, 1.17, and 1.10 \msun ~on the 1.5-Gyr isochrone.
\label{fig:cmd}
}
\end{figure}

\begin{figure}
\plotone{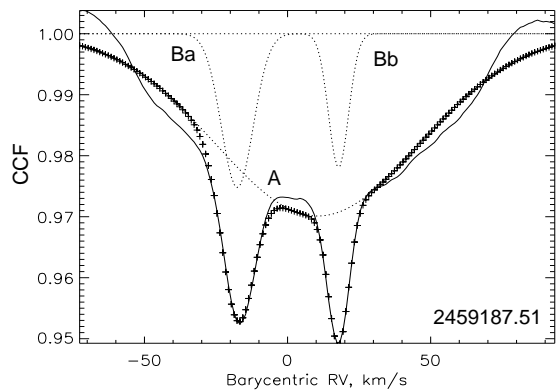}
\caption{  Cross-correlation function  (line) of  the spectrum  of ADS
  15988 taken on 2020 December 3 showing its triple-lined nature.  The
  dotted curves  are fitted Gaussians  and the crosses are  their sum.
  The  RVs of  Ba, Bb,  and  A are  $-17.56$, 17.81,  and 10.50  \kms,
  respectively.
\label{fig:CCF}
}
\end{figure}

\begin{deluxetable}{l c c cc  }
\tabletypesize{\scriptsize}     
\tablecaption{Parameters of Components of ADS 15988
\label{tab:STF2192} }  
\tablewidth{0pt}                                   
\tablehead{                                                                     
\colhead{Parameter} & 
\colhead{AB} &
\colhead{A} &
\colhead{Ba} &
\colhead{Bb} 
}
\startdata
$V$ (mag) & 5.51 & 5.75  & 7.84 & 8.22 \\
$V-K$ (mag) & 0.93 & 0.81: & 1.26: & 1.36: \\
${\cal M}$ (${\cal M}_\odot$) & 3.97 & 1.70 & 1.17 & 1.10 
\enddata
\end{deluxetable}

The   edge-on   orbit   of   A,B  (Figure~\ref{fig:A})   computed   by
\citet{Sod1999}  fits  the latest  measurements  rather  well and,  in
principle, does  not require revision. However, the  errors of orbital
elements are not  given in the above paper, and  the system of weights
is not  known. Here  the orbit is  re-fitted using all  available data
with appropriate weights and rejecting the outliers.  The orbit of A,B
serves as a  reference for determination of the  inner mass ratio from
the resolved  measurements of the subsystem.  Both  orbits were fitted
simultaneously by {\tt orbit4.pro} using only speckle measurements and
fixing  all outer  elements except  $T_0$,  $\Omega$, and  $i$ to  the
values   determined  from   the   historic  data   set.   The   result
(Figure~\ref{fig:wobble})  yields the  wobble  factor $f  = 0.46  \pm
0.02$, hence $q_{\rm Ba,Bb} = 0.83 \pm 0.07$.

Table~\ref{tab:STF2192} gives  the combined and  individual magnitudes
of ADS  15988 in the  $V$ and $K$  bands based on the  available data.
The  magnitude  difference  between  A  and B  has  been  measured  by
\citet{Davidson2009} ($\Delta V = 1.50 \pm 0.05$ mag, $\Delta R = 1.35
\pm 0.07$ mag), by \citet{Fab2000}  ($\Delta V =1.51$ mag, $\Delta B =
1.78$ mag), and  by Hipparcos ($\Delta Hp = 1.539$  mag).  It is clear
that B is slightly redder than A.  The mean of six measurements of the
magnitude difference between Ba and Bb  at SOAR in the $y$ band (close
to $V$) is 0.38 mag, with an rms scatter of 0.05 mag. The differential
photometry  and  the combined  $V_{\rm  AB}  =  5.51$ mag  define  the
components'   individual   magnitudes.    However,   no   differential
photometry in the  $K$ band is available, and  only the combined $V-K$
color is known.   This leaves two unknowns for  the three $V-K$ colors
of the components.  The colors of Ba and Bb in Table~\ref{tab:STF2192}
are  chosen   to  place  them  on   the  main  sequence   in  the  CMD
(Figure~\ref{fig:cmd}), thus also defining  the color of A. Obviously,
this is only a plausible guess, not a real measurement.

The most massive star A in  this system appears to be slightly evolved
off the  main sequence; it  is located on  the 1.5 Gyr  isochrone. The
masses of the  stars derived from this isochrone  and the absolute $V$
magnitudes     are     listed      in     the     last     line     of
Table~\ref{tab:STF2192}.  Adopting  these  masses, the  magnitudes  in
other  photometric bands  can be  deduced from  the  isochrone.  These
estimates lead to the magnitude difference between A and B of 1.74 and
1.38 mag in the $B$ and $R$ bands, respectively, in agreement with the
observed differential colors.

The masses of Ba and Bb estimated from the isochrone correspond to the
mass sum  of 2.27  \msun which  agrees with the  measured mass  sum of
2.34$\pm$0.25, within the error.  The isochrone masses imply the inner
mass ratio  of 0.94, somewhat  larger than 0.83$\pm$0.07  derived from
the wobble amplitude.  The total  mass sum deduced from the isochrone,
3.97 \msun, is slightly larger than measured (3.56$\pm$0.28 \msun) and
corresponds to the dynamical parallax of 18.4\,mas.

A    high-resolution    spectrum   taken    on    2020   December    3
(Figure~\ref{fig:CCF}) shows  lines of  all three components.  The RVs
match   qualitatively  their   expected  values   and   allow  correct
identification of the orbital nodes.  They are not used in the orbital
fit, pending additional observations. Future joint analysis of RVs and
resolved measurements will allow  us to measure masses without relying
on the isochrone and parallax.

The  mutual  inclination is  37\fdg5$\pm$2\fdg0.   Inclination of  the
outer  orbit, 89\fdg9$\pm$0\fdg06,  suggests possible  eclipses during
the conjunctions.  However,  star B is a close  pair, so only eclipses
involving its  components can take  place.  The next  conjunction will
happen around 2042.

\section{Summary}
\label{sec:sum}

In  most hierarchical  systems  studied here  the  outer binaries  are
bright  classical pairs with  ADS numbers  \citep{ADS} observed  for a
century  or  longer.  Their  inner  subsystems  are either  discovered
recently or also known from  the era of visual observations. The published
orbital elements are revised here using recent observations and appropriate
weights  and accounting for  the  wobble.  Some  long-period orbits  remain
poorly  constrained  even  when  the  expected mass  sum  is  used  as
additional input. Joint fitting of the inner and outer orbits 
 using all available information is the main result of this
paper.  The  orbits, together with  the estimated masses,  will enable
detailed   study   of  the   internal   dynamics   in  these   systems
\citep[e.g.][]{Xia2015,Hamers2020}.   The  ratios  of inner  and  outer
separations are moderate, so some interaction between the orbits is expected.

Our  small   and  random  sample   of  hierarchical  systems   is  not
statistically  relevant, but  it  does illustrate  the diversity of  their
architectures, ranging from  approximately aligned configurations with
small  eccentricities  to  highly  inclined or  even  counter-rotating
systems where  eccentric inner  orbits likely result  from the Kozai-Lidov
cycles.   This study  enlarges the  small collection  of  systems with
resolved  inner and outer  orbits. Previous  analysis of  such systems
\citep{moments} indicated that a mixture of hierarchies aligned within
70\degr  ~and those  with  uncorrelated orbits,  in  a 8:2  proportion,
matches the observed distribution of mutual inclinations $\Phi$.

\acknowledgements

Some  data  used here  were  obtained  at  the Southern  Astrophysical
Research (SOAR) telescope.  This work used the SIMBAD service operated
by Centre des Donn\'ees Stellaires (Strasbourg, France), bibliographic
references from  the Astrophysics Data System  maintained by SAO/NASA,
and the Washington  Double Star Catalog maintained at  USNO;  I thank
B.~Mason for  extracting historic  measurements from the  WDS database.
This  work has  made  use of  data from  the
European       Space       Agency       (ESA)       mission       Gaia
(\url{https://www.cosmos.esa.int/gaia}),  processed by  the  Gaia Data
Processing        and         Analysis        Consortium        (DPAC,
\url{https://www.cosmos.esa.int/web/gaia/dpac/consortium}).     Funding
for the DPAC has been provided by national institutions, in particular
the institutions participating in the Gaia Multilateral
Agreement. Work of the author is funded by the NSF's NOIRLab.

\facility{Facility: SOAR}

\end{document}